\documentclass[journal,12pt,onecolumn,letterpaper]{IEEEtran}
\usepackage{arxiv}
\usepackage{geometry}
\usepackage{times}
\usepackage{cite}
\usepackage{url}
\usepackage{graphicx}
\usepackage{lscape}
\usepackage{subcaption}
\usepackage{rotating}
\usepackage{rotfloat}
\usepackage{xcolor}
\usepackage{amsmath}
\usepackage{amssymb}
\usepackage[linesnumbered,ruled,vlined]{algorithm2e}
\usepackage{pseudocode}
\usepackage{array}
\usepackage[english]{babel}
\usepackage{gensymb}
\usepackage{textcomp}
\usepackage{placeins}
\usepackage{balance}
\usepackage{booktabs}

\title{CAHICHA: Computer Automated Hardware Interaction test to tell Computer and Humans Apart \\ 
}

\author{%
  
    Aditya Mitra \\
    Centre of Excellence, Artificial Intelligence \& Robotics (AIR),\\
    School of Computer Science and Engineering\\
    VIT-AP University, India \\
    DigitalFortress Private Limited \\
    \texttt{adityaarghya0@gmail.com}
\And

  Sibi Chakkaravarthy Sethuraman\\
    Centre of Excellence, Artificial Intelligence \& Robotics (AIR),\\
    School of Computer Science and Engineering\\
    VIT-AP University, India \\
    DigitalFortress Private Limited \\
    \texttt{sb.sibi@gmail.com} \\
\And
  Devi Priya V S \\
    Department of CS (Cyber Security), \\
    School of Engineering, \\
    Dayananda Sagar University, \\
     Harohalli,Karnataka, India \\
    \texttt{vsdevipriya@gmail.com} \\
}

\begin{document}
\maketitle


\begin{abstract}
As automation bot technology and Artificial Intelligence is evolving rapidly, conventional human verification techniques like voice CAPTCHAs and knowledge-based authentication are becoming less effective. Bots and scrapers with Artificial Intelligence (AI) capabilities can now detect and solve visual challenges, emulate human-like typing patterns, and avoid most security tests, leading to high-volume threats like credential stuffing, account abuse, ad fraud, and automated scalping. This leaves a vital gap in identifying real human users versus advanced bots. We present a novel technique for distinguishing real human users based on hardware interaction signals to address this issue. In contrast to conventional approaches, our method leverages human interactions and a cryptographically attested User Presence (UP) flag from trusted hardware to verify genuine physical user engagement—providing a secure and reliable way to distinguish authentic users from automated bots or scripted routines. The suggested approach was thoroughly assessed in terms of performance, usability, and security. The system demonstrated consistent throughput and zero request failures under prolonged concurrent user demand, indicating good operational reliability, efficient load handling, and the underlying architecture's robustness. These thorough analyses support the conclusion that the suggested system provides a safer, more effective, and easier-to-use substitute for current human verification methods.

\end{abstract}

\maketitle
%
%


\section*{Introduction}
AI is not only enabling attackers to develop increasingly sophisticated, elusive bots that target Application Programming Interface (APIs), take advantage of business logic, and commit fraud, but also making it easier for attackers to gain access to the cyber domain, which leads to a rise in the number of basic bot attacks. Surpassing human activity for the first time in ten years, bots account for 51\% of all online traffic, with over two-thirds of that traffic being malicious in 2025 compared to 42\% in 2024 \cite{Pressrelease,Pressrelease2025}. The rise in Distributed Denial of Service (DDoS) attacks and botnet-driven operations have placed companies at significant risk \cite{wang2018delving}. A botnet is a dispersed network of endpoints (bots or zombies) contaminated with malware that is managed by a command-and-control (C2) system \cite{tariq2025alert,hoque2015botnet}. These compromised systems are coordinated by the botmaster to produce large amounts of malicious traffic, such as SYN floods \cite{alarnaout2022rapt}, HTTP GET/POST floods \cite{do2021methodology}, or amplification-based requests \cite{griffioen2021scan}, directed to the victim system in a DDoS attack. Botnet-driven DDoS attacks are extremely scalable, durable, and challenging to counteract due to their dispersed and volumetric nature \cite{ debicha2023adv}. 

As websites require a way to distinguish between automated bots that engage in illicit activities ( creating fictitious accounts, stuffing credentials, scraping data, and attempting DoS attacks), Completely Automated Public Turing test to tell Computers and Humans Apart (CAPTCHAs) are extensively utilized. Various research works \cite{guerar2021gotta,wang2023experimental}  have critically assessed CAPTCHA mechanisms based on their security weaknesses and usability issues. Classic text-based CAPTCHAs previously thought of being strong were made useless against sophisticated Optical Character Recognition (OCR) technologies \cite{zhang2020robust} and machine learning (ML) mechanisms \cite{dionysiou2020sok}. Studies further demonstrated that contemporary artificial Intelligence (AI) models can attain nearly ideal rates of circumventing reCAPTCHA by emulating human-like mouse movement \cite{shahania2025gotta}.
As computing power and AI systems such as massive language models and powerful computer vision increases, automated systems are becoming more adept at solving captchas, which were designed to distinguish them from the website's genuine users. This has led to considerable web-scraping for training AI models, as it can now automatically access the illicit domains. Moreover, this increases the possibility of DoS  and DDoS assaults caused by excessive automated traffic that can bypass the CAPTCHA.
 The designers of captcha systems have subsequently resorted rendering captchas even more challenging in order to prevent automation, but this has mainly turned out to be counterintuitive; it has also become very confusing, time-consuming, and painful for human users. However, AI models can solve the more recent CAPTCHAs since they can learn and train more quickly than the ordinary human \cite{alajmi2020password}.

AI-driven scraping and crawling have developed far beyond mere rule-based bots to being adaptable smart agents that can simulate human browsing patterns \cite{khder2021web,murty2022building}. Their valid applications involve bulk data gathering for e-commerce, price tracking, financial market forecasting, digital marketing analytics, scholarly research, and cyber threat intelligence gathering. In contrast to conventional crawlers, artificial intelligence-based systems utilize reinforcement learning, machine learning, and natural language processing (NLP) technique to dynamically understand website structures, evade CAPTCHAs, and adapt to real-time anti-bot policies. All these developments, however, pose extreme threats to the safety of browsers and web infrastructure security. For example, bad actors can weaponize AI scrapers to perform reconnaissance on websites, scraping sensitive metadata like API endpoints, session tokens, or configuration files. In addition, attackers can stage mass-scale, distributed AI crawlers that create adaptive, human-like traffic patterns that web servers and and other backend processes equally — thus spreading into DDoS states that bypass traditional detection thresholds. From the point of view of browser security, the problem is exacerbated since classical defenses (e.g., rate limiting, static CAPTCHAs, user-agent checks, and cookie tracking) are becoming less effective against AI-powered crawlers. Such agents are able to mimic legitimate session flows, handle dynamic cookies, rotate IPs automatically, and even mimic keystroke dynamics in order to avoid fingerprinting. This undermines the effectiveness of existing browser based security controls, placing users and organizations at risk from covert scraping, session hijack attempts, and secondary DDoS amplification. In addition to changing user interactions, the advent of sophisticated AI tools such as ChatGPT, ByteSpider Bot, ClaudeBot, Google Gemini, Perplexity AI, and Cohere AI is also changing how attackers carry out cyberthreats. The Imperva Threat Research team reports \cite{Thales_2025} that 54\% of all AI-enabled assaults are caused by the ByteSpider Bot alone, demonstrating how commonly used AI technologies are being utilized for cyberattacks. Other notable contributors are ChatGPT User Bot (6\%), ClaudeBot (13\%), and AppleBot (26\%).

Browser-based investigations check the JavaScript code, user agents, cookies, and occasionally initiate invisible CAPTCHA-like challenges \cite{nayak2024experimental}. Additionally, they examine behavioral indicators such as typing speed, network traffic, fingerprinting information, and mouse movements. Furthermore, for server-side verification to work, the browser must correctly answer cryptographic challenges. This guarantees that it is a legitimate browser and not a script without a head. However, a significant drawback of these techniques is that all verification takes place inside the browser. Sometimes, sophisticated bots  such as headless browsers, automation tools like Selenium and Puppeteer, or AI-driven scripts \cite{golian2025integrated} can circumvent all these measures.
 To address all these, our research proposes a novel approach that utilize hardware interaction of the end user which is verifiable with cryptographic attestation. The hardware-rooted cryptographic attestation ensures a real physical interaction occurred, making automated attacks costly, complex, and difficult to scale.
 
The significance of our Experimentation :

\begin{itemize}

\item Our approach uses direct hardware interaction and provides a more reliable and impenetrable means of differentiating humans from automated systems considering the fact that software bots cannot be programmed to communicate with sensors, operate buttons, or input biometric information.
\item With this approach, adversaries are forced to rely on expensive, non-scalable robotic hardware setups as it enforces cryptographically attested physical events, unlike browser-based defenses that can be circumvented by headless automation frameworks imitating execution environments.
\item It ensures that only genuine physical events from trusted hardware are accepted thus eliminates the risk of software-based emulation or spoofing.

\end{itemize}


\section*{Background and Related Work}
\begin{table}[]
\caption{Comparative Study of Existing CAPTCHA Methods}
\label{Comparative Study of Existing CAPTCHA Methods}
\begin{tabular}{|l|l|l|l|}
\hline
\multicolumn{1}{|c|}{\textbf{Type}} & \multicolumn{1}{c|}{\textbf{How It Works}} & \multicolumn{1}{c|}{\textbf{Advantages}} & \multicolumn{1}{c|}{\textbf{Limitations}} \\ \hline
\textbf{Text-based \cite{zhang2022counteracting}} & \begin{tabular}[c]{@{}l@{}}User types distorted characters\\ displayed in an image\end{tabular} & \begin{tabular}[c]{@{}l@{}}Simple to implement, \\ widely supported\end{tabular} & \begin{tabular}[c]{@{}l@{}}Bots using OCR can bypass; \\ difficult for visually impaired\\  users\end{tabular} \\ \hline
\textbf{Image-based \cite{gutub2023practicality} } & \begin{tabular}[c]{@{}l@{}}User selects images matching a\\ theme (e.g., “Select all cars”)\end{tabular} & \begin{tabular}[c]{@{}l@{}}Harder for bots than \\ text; interactive\end{tabular} & \begin{tabular}[c]{@{}l@{}}Time-consuming for users; \\ AI image recognition may \\ bypass\end{tabular} \\ \hline
\textbf{Audio \cite{wang2023improving}} & \begin{tabular}[c]{@{}l@{}}User listens to distorted speech \\ and types it\end{tabular} & \begin{tabular}[c]{@{}l@{}}Accessible for visually \\ impaired users\end{tabular} & \begin{tabular}[c]{@{}l@{}}Background noise can make it\\  hard; vulnerable to speech \\ recognition bots\end{tabular} \\ \hline

\textbf{reCAPTCHA (Google) \cite{avanesi2022m}} & \begin{tabular}[c]{@{}l@{}}Advanced risk analysis, image \\ recognition, or checkbox \\ (“I’m not a robot”)\end{tabular} & \begin{tabular}[c]{@{}l@{}}Easy for humans, \\ integrates ML\end{tabular} & \begin{tabular}[c]{@{}l@{}}Privacy concerns; bots can \\ bypass with AI; may fail for \\ non-tech-savvy users\end{tabular} \\ \hline
\textbf{Mathematical \cite{guerar2021gotta}} & Solve a simple math problem & \begin{tabular}[c]{@{}l@{}}Lightweight, easy to \\ implement\end{tabular} & \begin{tabular}[c]{@{}l@{}}Easily solved by bots with \\ basic algorithms; poor UX for \\ complex questions\end{tabular} \\ \hline
\textbf{Behavioral \cite{gangwal2025swiss} } & \begin{tabular}[c]{@{}l@{}}Detects human behavior like\\ mouse movement or scrolling \\ patterns\end{tabular} & \begin{tabular}[c]{@{}l@{}}Transparent to users; \\ user-friendly\end{tabular} & \begin{tabular}[c]{@{}l@{}}Can be bypassed by highly
\\sophisticated bots; requires \\more backend processing\end{tabular} \\ \hline
\textbf{Slide/Drag \cite{teoh2025captchas} } & \begin{tabular}[c]{@{}l@{}}User drags a slider to complete \\ an action\end{tabular} & Interactive and simple & \begin{tabular}[c]{@{}l@{}}Can be automated by advanced \\ scripts; not always mobile-\\friendly\end{tabular} \\ \hline
\end{tabular}
\end{table}
 Techniques for CAPTCHA have advanced dramatically in order to strike a compromise between user experience and security. Conventional text-based CAPTCHAs employ distorted characters to thwart automated detection.  Puzzle and interactive CAPTCHAs leverage human dexterity and intelligence to engage users in activities like dragging or rotating items. Object recognition tests and other image-based CAPTCHAs are more user-friendly and more difficult for bots to complete without sophisticated computer vision capabilities. Though they have drawbacks with speech recognition software and user comprehension, audio CAPTCHAs provide accessibility by requiring users to transcribe spoken content. As laid out by invisible CAPTCHA systems like reCAPTCHA v3, behavioral CAPTCHAs leverage user activities, such as mouse movements and scrolling patterns, to passively identify bots. A comparative study of prominent state- of-the-art CAPTCHA methods is summarized in table \ref{Comparative Study of Existing CAPTCHA Methods}.  The following sections provides an overview of the technology employed in this work.

\subsection*{FIDO2 (Fast IDentity Online)}

The terms FIDO2 (Fast IDentity Online) and WebAuthn are frequently used interchangeably. Despite their close relationship, they are not exactly the same. FIDO2 is an open standard created by the FIDO Alliance that allows users to access desktop and mobile applications without a password \cite{angelogianni2024many}. 
FIDO2 is a combination  of two complementing parts: (a) WebAuthn, that allows applications to employ cryptographic possession-based authentication to verify users. (b) Client To Authentication Protocol (CTAP), which allows the client to connect to a roaming authenticator, like a smartphone or a hardware security key \cite{ghosh2024saila}. 

\subsection*{Web Authentication (WebAuthn)}
WebAuthn is a web standard for browsers that was released by the World Wide Web Consortium (W3C) \cite{vasileios2021web,ghosh2024saila}. In contrast to knowledge-based credentials, it provides a controlled secure method for users to login in conjunction with a possession component such as hardware security keys or special cryptographic chips in the user device. WebAuthn replaces or improves the authentication component of many websites that currently include pages that let users create new accounts or log in to existing ones. By abstracting user agent-authenticator communication, it expands the Credential Management API \cite{stebila2024quantum}. With hardware-backed authenticators like security keys, or trusted platform modules (TPMs), it enables a web application to register and authenticate users. WebAuthn guarantees that only legitimate human interactions from registered authenticators are recognized by providing cryptographically verifiable proof of user presence (UP) and user verification (UV). These attestation signals provide a strong layer of bot mitigation by being able to distinguish between automated scripts and humans in addition to standard authentication. Public key cryptography is utilized by WebAuthn to provide safe, passwordless authentication with the UP and UV flags to cryptographically attest the user presence and his intent. 

\subsection*{Client To Authenticator Protocol (CTAP)}

The specification of Client To Authenticator Protocol (CTAP) outlines how an operating system and application (such as a browser) communicate with a compatible authenticator[device] via Bluetooth, Near-Field Communication (NFC), or Universal Serial Bus (USB) \cite{kuchhal2023evaluating}. Secure communication between a user device and an authenticator during the registration and authentication process is facilitated by CTAP. Through a defined transit channel, the client sends the request to the authenticator whenever an application initiates the authentication or enrollment process. Following user authentication or enrollment, the authenticator creates cryptographic replies and safely sends the results back to the client. This response is sent to the relying party by the client, who completes the authentication or enrollment process and verifies the proof \cite{bindel2023fido2}.

\section*{ Proposed Methodology}

Without modifying the internals, the suggested solution can be implemented as a proxy server in the vicinity of an already-existing website or web application.

\subsection*{Technical Overview }\label{sec4}
\begin{figure}[h!]
\centerline{\includegraphics[width=12cm,height=7cm]{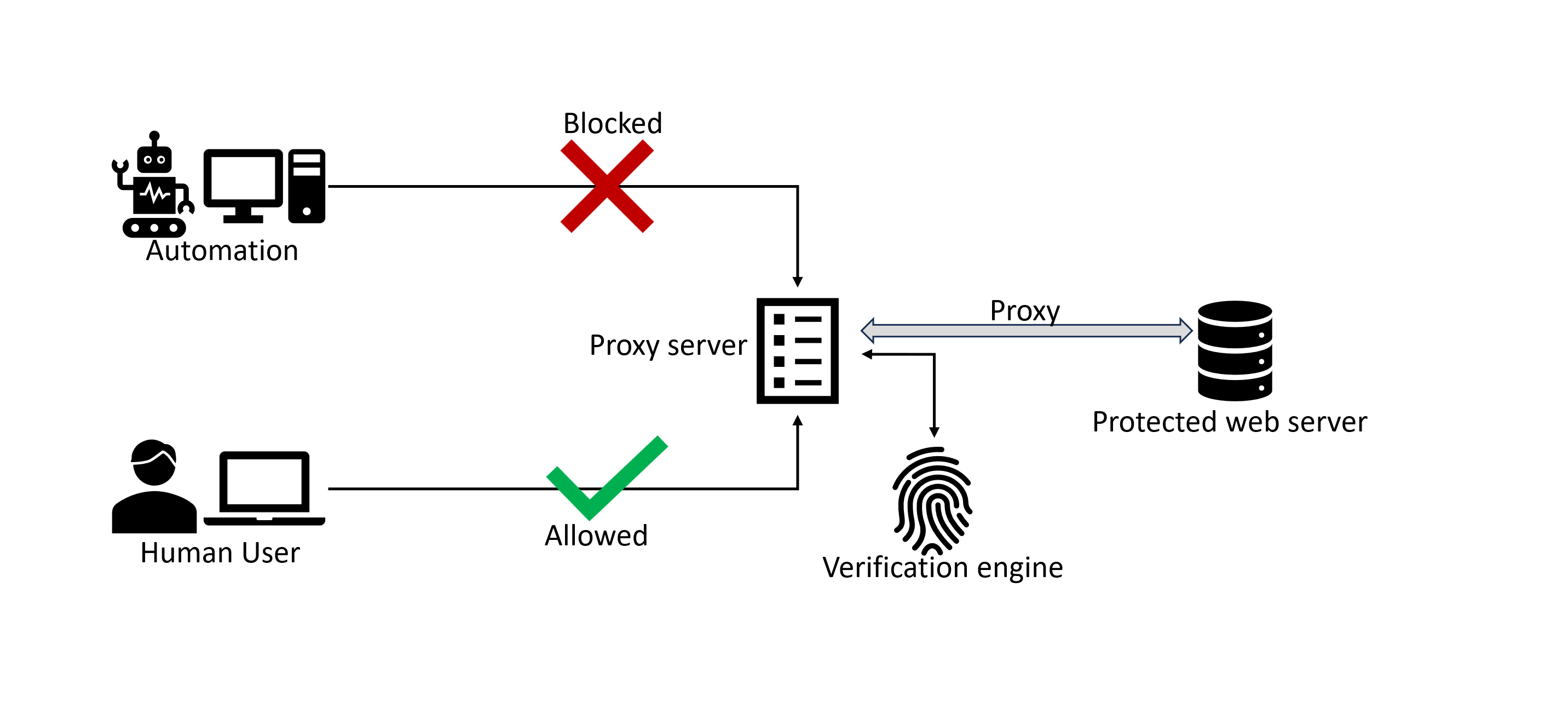}}
\caption{Workflow of CAHICHA}
\label{architecture}
\end{figure}

Although FIDO2 and Web Authentication are commonly utilized as phishing-resistant authentication solutions, our solution examines them from a different perspective by employing them to distinguish between automation and people using the User Verification (UV) and User Presence (UP) flags. The deployment architecture is depicted in Figure \ref{architecture}.
The proxy server verifies the existence and authenticity of a verification cookie when a user visits the website. The user clearly does not have the cookie information when they connect for the first time. The verification engine now creates public key credential creation options with the User Presence and User Verification fields designated as “required” and sends it to the client browser. Then the user's browser subsequently calls the Navigator.credentials.create()  function \& generate a FIDO2 credential \cite{mahdad2024breaching}. The origin (the URL host) and the challenge are included in the Client Data that the browser generates. The authenticator, which could be a physical security key or the user computer's Trusted Platform Module (TPM), receives this hashed client data. The system then introduces a process in which the user must interact with hardware physically (cryptographic attestation task). By requiring the user to complete a verification, such as entering a PIN via a secure channel, enrolling in a biometric modality (such a fingerprint sensor), or physically touching a touch sensor on the FIDO2 hardware security key, the relying party solicits the user to verify. Physical interaction with a hardware device verifies that these processes are bound to secure elements and Trusted Execution Environments (TEE). It is therefore impossible for malicious scripts or headless processes to automate these procedures, even if the attacker uses an automated mechanism to trick the browser. During attestation, the verification response—a biometric match, PIN entry, or metal contact—is confirmed as cryptographically associated to the authenticator device. Once the user verification step is completed, the authenticator will return the attestation response. The authenticator will make a new public–private key pair and build the authenticator data structure. The authenticator data has the following fields;
\begin{itemize}
    \item RP ID Hash (32 bytes): The SHA-256 hash of the Relying Party Identifier (RP ID) which is the host part of the web origin or a valid registrable parent domain.
\item Flags (1 byte): A bitfield with several bits for different status indicators, the most important bits in consideration being:
\begin{itemize}
    \item Bit 0 (UP – User Presence): Set if an action has been taken on the authenticator by the user (e.g. button press, touch). This is a protection from automated user interaction.
    \item Bit 2 (UV – User Verification): Set if the user successfully completed a local verification (e.g. pin, biometric).
\end{itemize}

\item 
Signature Counter (4 bytes): A monotonically increasing counter used to detect cloned authenticators.
\item Attested Credential Data (variable length): This field consists of metadata regarding the credential and the public key associated with the just created key pair.
\end{itemize}

\begin{figure}[h!]
\centerline{\includegraphics[width=17cm,height=9cm]{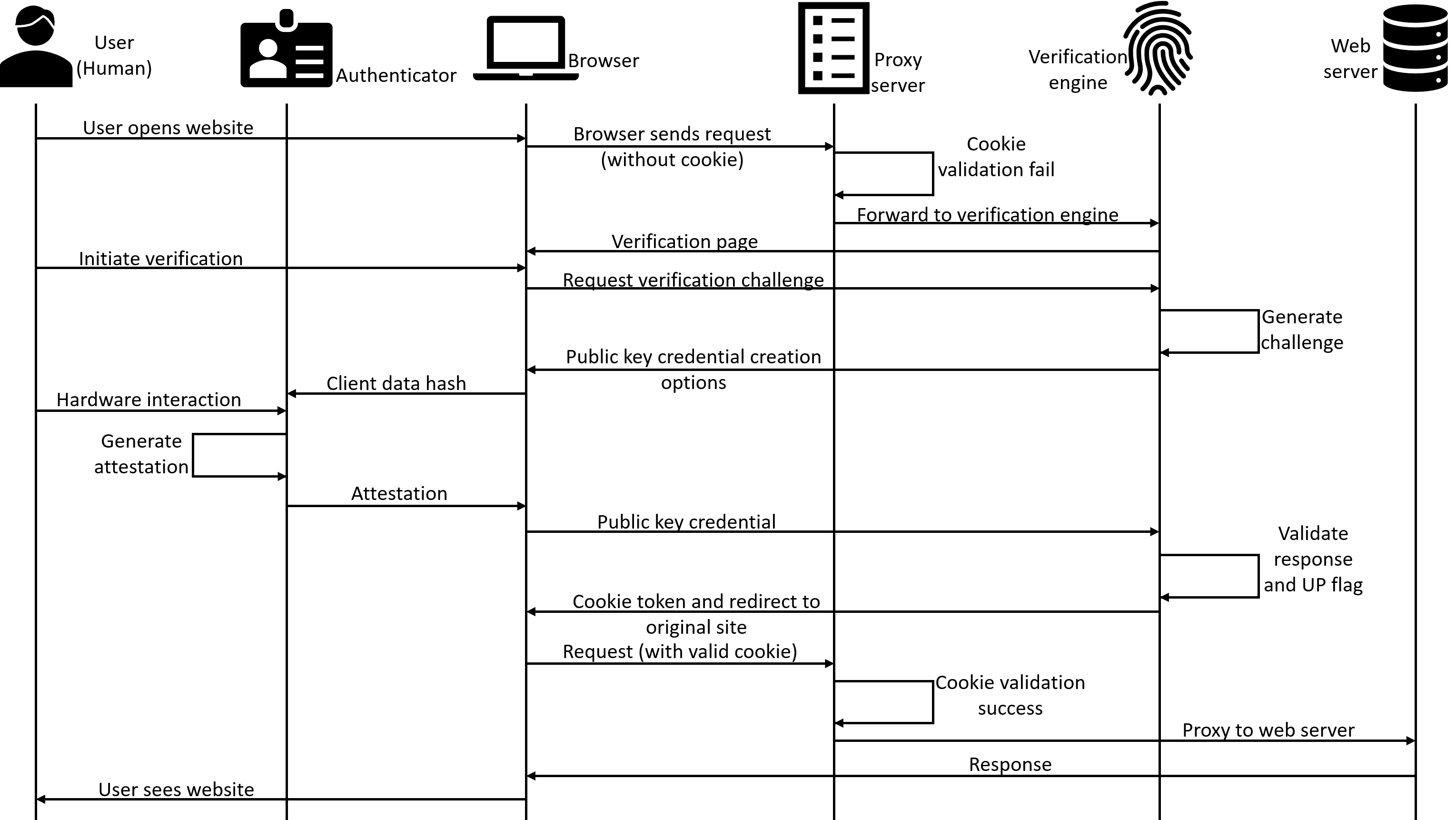}}
\caption{Sequential Diagram of CAHICHA}
\label{Sequential Diagram1}
\end{figure}

Once the authenticator data (authData) has been created, the authenticator concatenates it with the SHA-256 hash of the client data (clientDataHash). This byte string is signed with a private key from the newly generated credential key pair for matching with the corresponding public key in the attested credential data so that the RP can verify the signature. Also, and depending on the authenticator’s make and model, the whole attestation statement might also be signed using an attestation private key that was given or provisioned to the authenticator by the manufacturer of the authenticator in the manufacturing process, providing a way to obtain cryptographic proof of the device’s provenance. For certified authenticators, the certificate chain of the manufacturer’s attestation certificate is embedded in the attestation statement, and validities can be vetted against the FIDO Alliance Metadata Service (MDS v3), allowing RPs to verify that the authenticator is legitimate and provides security assurances that it has not been tampered with. The attestation response is sent to the Verification Engine which will perform the following checks:
\begin{itemize}
    \item Signature Verification:
The engine validates the digital signature over the concatenation of authenticatorData || clientDataHash using the public key from the attested credential data. This ensures integrity and authenticity of the response. 
\item 
Challenge binding and replay protection : The engine will compare the challenge embedded in the clientDataJSON with the original challenge that was generated and held by the RP. If two challenges do not match, that means the response was either replayed from a prior session or tampered with, therefore providing replay protection.
\item Authenticator Data Evaluation: The Flags field in the authenticator data has been parsed. UP (User Presence) bit (bit 0) should be set, as a guarantee that there was some trusted hardware mediated user interaction (touch or button push), ensuring credential creation was initiated by the user and not entirely automated.
\item Authenticator Trustworthiness :
When in strict attestation validation mode (i.e., CAHICHA strict mode:-To make sure that only legitimate hardware-based security keys are used for authentication, it implements FIDO Metadata Service (MDS) verification. ), the Verification Engine will verify the attestation certificate chain present in the response. 
The certificate chain will be validated against trusted roots present in the FIDO Alliance Metadata Service (MDS v3).
This will declare that the authenticator that issued the response was a certified device meeting the respective security requirements of FIDO, not a fake or software authenticator. It is to be noted that, though MDS3 confirms the legitimacy of a security key, there are various keys available commercially which are not enrolled in the MDS3. The system would not be able to verify these keys. Hence the system offers two deployment modes: Strict mode and General mode. In Strict mode, the MDS3 verification is enforced and is used in cases where all or most users are sure to have a FIDO Alliance certified authenticator device (like in enterprise scenarios). General mode, on the other hand, allows using any authenticator device and this may allow attackers to use devices modified at hardware level to bypass CAHICHA
\end{itemize}

Once the Verification Engine establishes that the request originates from a legitimate human user, it issues a session token.
\begin{itemize}
    \item Token Generation:
The token is built as a structured payload with a high-resolution timestamp (the UNIX epoch in milliseconds) and a fixed identifier string (protocol magic value). The Fernet cryptography library, which uses a 128-bit key and the Advanced Encryption Standard (AES) in Cipher Block Chaining (CBC) mode internally, is then used to encrypt this payload. By integrating base64 encoding, integrity verification, and encryption into a single, user-friendly format, Fernet offers a higher-level abstraction than AES-CBC.

\item Client Storage:
The encrypted token is issued to the client as an HTTP cookie. 
The cookie includes attributes such as HttpOnly, Secure, etc \cite{kwon2019security}. 
The client is then redirected back to the original resource endpoint.
\item Proxy Server Validation:
Upon receiving subsequent client requests, the Proxy Server retrieves the cookie and decrypts the token using the shared symmetric key. It verifies- a)the presence of the fixed identifier string. b) Timestamp validity: if the token age exceeds 24 hours, it is rejected. 
If validation fails, the Verification Engine challenge is re-initiated.

\item Request Forwarding: 
If the token is valid, the Proxy Server transparently forwards the request to the origin web server.
\end{itemize}
The sequential diagram of the CAHICHA system, which shows the successive interactions between its components, is shown in Figure \ref{Sequential Diagram1}.

\section*{Results and Analysis}
\begin{figure}[h!]
\centerline{\includegraphics[width=15cm,height=8cm]{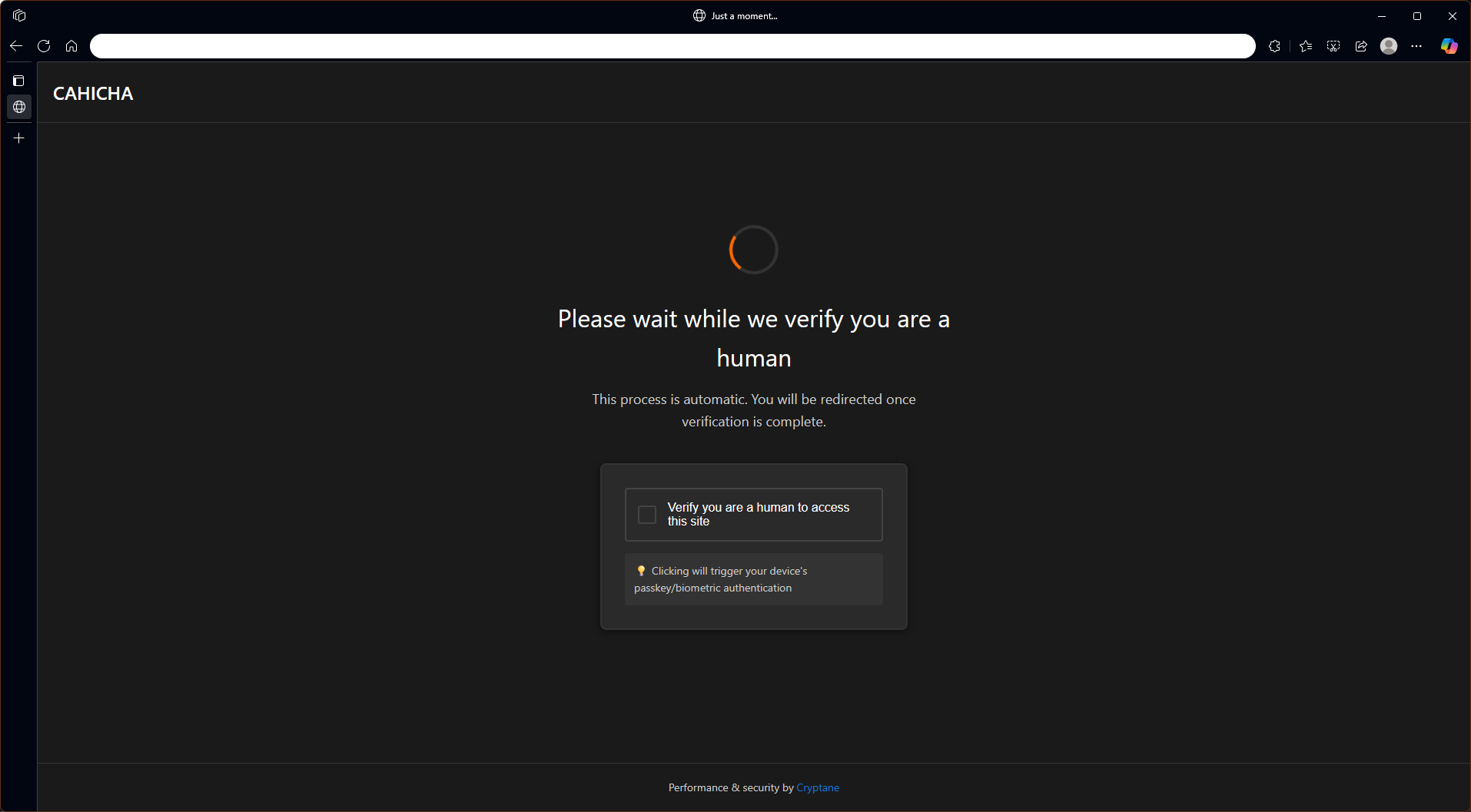}}
\caption{CAHICHA Login Verification Interface}
\label{CAHICHA Login Verification Interface}
\end{figure}

\begin{figure}[h!]
\centerline{\includegraphics[width=15cm,height=8cm]{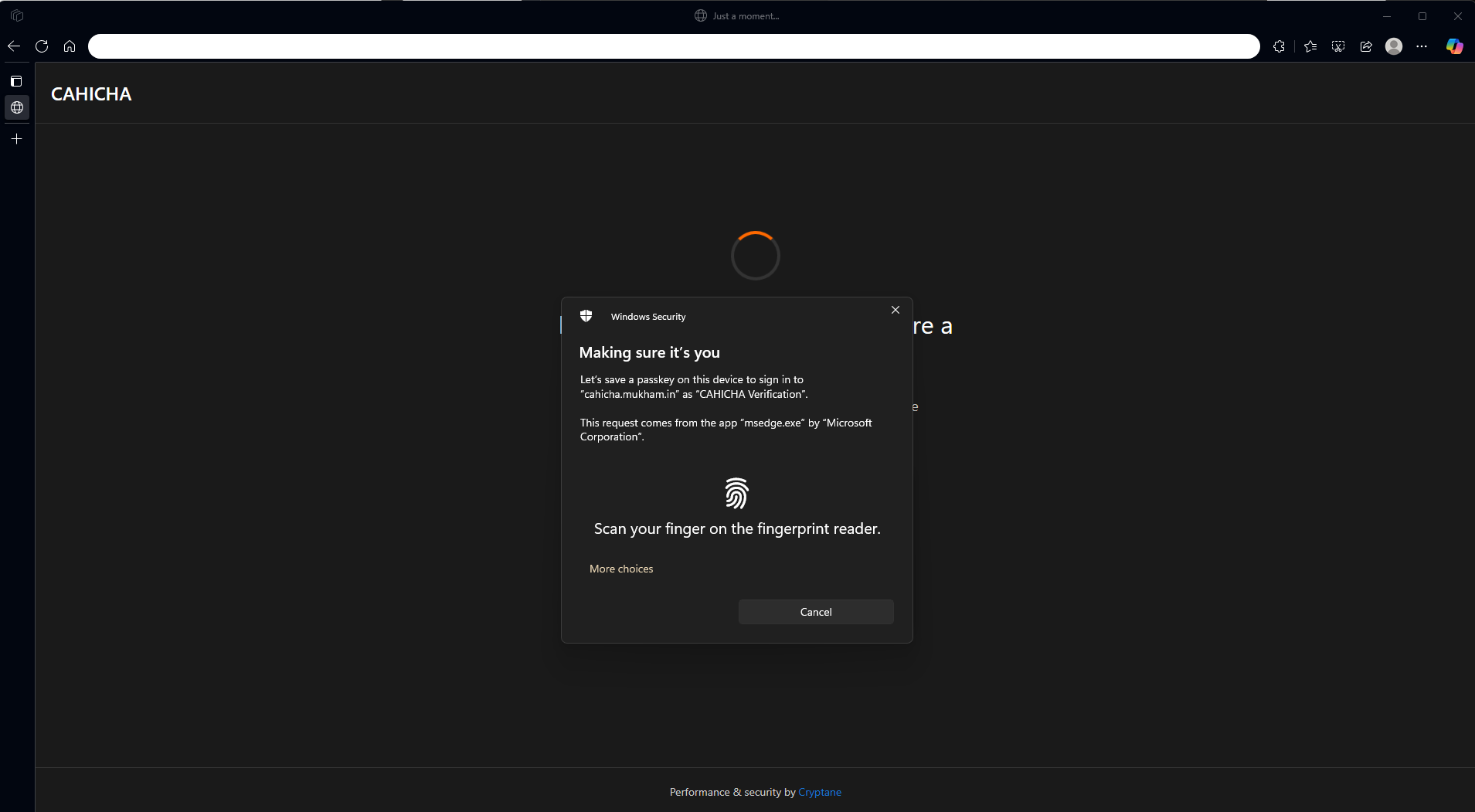}}
\caption{Biometric Verification for Human Authentication}
\label{Biometric Verification for Human Authentication}
\end{figure}

Within the test environment, a demonstration web app is installed on Virtual Private Cloud (VPC) instance of SSD Nodes, a cloud service provider (CSP). The backend of the application is served via the Apache2 web server and is set to operate over TCP port 8080. For hardening purposes, incoming access to port 8080 is specifically limited at the host level through Uncomplicated Firewall (UFW) rules so that the service cannot be accessed directly from the outside network. In addition, it ensures that all traffic ingress must pass through a proxy rather than directly accessing port 8080. To address automated scraping and access attempts made through bots, CAHICHA Apache2 plugin is incorporated as an application-layer verification mechanism. The plugin is set up with a reverse proxy towards the backend service that listens on port 8080 and hence imposes the condition that all requests are first checked by CAHICHA and then proxied to the guarded application. This architecture ensures that only validated human users have access, while AI-based bots and automatic crawlers are blocked at the proxy level. In addition, Transport Layer Security (TLS) is activated on Apache2 reverse proxy to ensure encrypted client–server communication, maintain data integrity, and guard user interactions from tampering or eavesdropping. This is also required since WebAuthn works only in a secure context. This multi-layered setup offers a managed test environment in which resilience can be assessed against automated crawling, scraping, and DDoS-like behavior within a browser-confronting environment.

In Figure \ref{CAHICHA Login Verification Interface}, the user is prompted to authenticate rather than using conventional CAPTCHA testing (when the user has already completed a verification activity). The methodology employs device-level hardware interaction. At this point, the operating system and the browser request that the user scan their fingerprint.
Figure \ref{Biometric Verification for Human Authentication} illustrates a CAHICHA-activated biometric verification dialog window that requires the user to authenticate with a finger scan. However, depending on the physical security key setup, the system may just need the user to touch a touch sensor only. This hardware interaction verification utilizes a cryptographic device to attest the hardware interaction. Hardware-based verification is resistant to automated bot attacks, credential theft, and replay attempts, making it more secure than traditional CAPTCHA modalities. A feature-based comparitive analysis between CAHICHA (in both the strict and General modes) and traditional CAPTCHA systems, such as reCaptcha v3 and hCaptcha, is shown in Table \ref{Feature-Based Comparison of CAHICHA and Conventional CAPTCHA Systems}. 

\begin{table}[!htbp]

\caption{Feature-Based Comparison of CAHICHA and Conventional CAPTCHA Systems}
\label{Feature-Based Comparison of CAHICHA and Conventional CAPTCHA Systems}
\begin{tabular}{|l|l|l|l|l|}
\hline
\textbf{Feature} & \textbf{\begin{tabular}[c]{@{}l@{}}CAHICHA \\ (Strict mode)\end{tabular}} & \textbf{\begin{tabular}[c]{@{}l@{}}CAHICHA \\ (General mode)\end{tabular}} & \textbf{ReCaptcha v3} & \textbf{hCaptcha} \\ \hline
\begin{tabular}[c]{@{}l@{}}Method of \\ usage\end{tabular} & \begin{tabular}[c]{@{}l@{}}Hardware Interaction, \\ validated \\ with   FIDO MDS\end{tabular} & Hardware Interaction & Behavioral Analysis & \begin{tabular}[c]{@{}l@{}}Image classification \\ and puzzles (Turing \\ Test)\end{tabular} \\ \hline
Speed & \begin{tabular}[c]{@{}l@{}}Fastest (Single touch \\ on a hardware).\end{tabular} & \begin{tabular}[c]{@{}l@{}}Fastest (Single touch \\ on a hardware).\end{tabular} & \begin{tabular}[c]{@{}l@{}}Slower (Behavioral \\ analysis and may\\ present puzzle if \\ score is low).\end{tabular} & Slowest. \\ \hline
\begin{tabular}[c]{@{}l@{}}Time taken \\ (Average)\end{tabular} & 12ms & 12ms & \begin{tabular}[c]{@{}l@{}}4s if score is high, \\ 10s if low\end{tabular} & 19s \\ \hline
\begin{tabular}[c]{@{}l@{}}User \\ Experience\end{tabular} & Non intrusive & Non intrusive & \begin{tabular}[c]{@{}l@{}}Non intrusive in \\ most cases \\ but   slower\end{tabular} & Very intrusive \\ \hline
Bot Detection & \begin{tabular}[c]{@{}l@{}}Deterministic hardware \\ based\end{tabular} & \begin{tabular}[c]{@{}l@{}}Deterministic hardware \\ based\end{tabular} & \begin{tabular}[c]{@{}l@{}}May be spoofed by \\ sophisticated \\ AI   models\end{tabular} & \begin{tabular}[c]{@{}l@{}}May be spoofed by \\ AI\end{tabular} \\ \hline
False positives & Zero & \begin{tabular}[c]{@{}l@{}}Edge cases where fake \\ authenticator is used \\ may result in false\\  positive\end{tabular} & \begin{tabular}[c]{@{}l@{}}High with \\ sophisticated \\ AI models and \\ browser\\  automation\end{tabular} & \begin{tabular}[c]{@{}l@{}}Highest with AI \\ models\end{tabular} \\ \hline
\begin{tabular}[c]{@{}l@{}}False \\ negatives\end{tabular} & \begin{tabular}[c]{@{}l@{}}Certain platform \\ authenticators \\ like Windows Hello \\ might result in false\\  negatives\end{tabular} & \begin{tabular}[c]{@{}l@{}}Zero (on supported \\ hardware)\end{tabular} & \begin{tabular}[c]{@{}l@{}}Users with vision \\ or cognitive \\ disorders might fail\end{tabular} & \begin{tabular}[c]{@{}l@{}}Users with vision or \\ cognitive   \\ disorders might fail\end{tabular} \\ \hline
Security & Highest & High & Low & Lowest \\ \hline
Setup & \begin{tabular}[c]{@{}l@{}}Easiest (Drop in as \\ a proxy without \\ change in website \\ code)\end{tabular} & \begin{tabular}[c]{@{}l@{}}Easiest (Drop in as a \\ proxy without change \\ in website code)\end{tabular} & \begin{tabular}[c]{@{}l@{}}Requires \\ website code \\ changes\end{tabular} & \begin{tabular}[c]{@{}l@{}}Requires widget \\ added on all \\ pages by \\ code change\end{tabular} \\ \hline
Accessibility & Excellent & Excellent & Poor & Poorest \\ \hline
\begin{tabular}[c]{@{}l@{}}Reliance on \\ 3rd party\end{tabular} & \begin{tabular}[c]{@{}l@{}}Minimal (self hosted). \\ Access to FIDO\\ Metadata services \\ required\end{tabular} & Zero (self hosted) & \begin{tabular}[c]{@{}l@{}}High (Google \\ services)\end{tabular} & High (hCaptcha) \\ \hline
Privacy & Highest (no tracking) & Highest (no tracking) & \begin{tabular}[c]{@{}l@{}}Low (Browser \\ activity \\ tracking)\end{tabular} & \begin{tabular}[c]{@{}l@{}}Lowest (hCaptcha \\ business model \\ sells labelled \\ data to AI/ML \\ companies)\end{tabular} \\ \hline
\end{tabular}
\end{table}

\subsection*{{Experimental Validation}}

\subsubsection*{ Stress/ Load Testing}

\begin{figure}[h!]
\centering
\includegraphics[width=17cm,height=13cm]{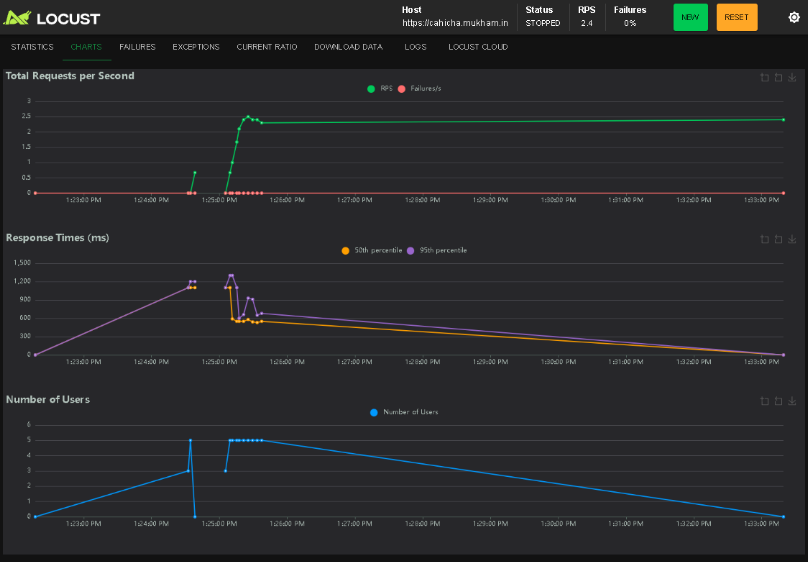}
\caption{Load Test Result using Locust}
\label{load test1}
\end{figure}

The goal of load testing is to determine how well a website functions at typical or projected traffic levels. Our objective is to guarantee that the system, especially the CAHICHA proxy can manage the usual volume of users and transactions without experiencing any deterioration in performance. The outcome of the experimentation to achieve this is shown in Figure \ref{load test1}.

Three important performance metrics of a web application or service are depicted in Figure \ref{load test1} over a ten-minute window from 1:25 PM to 1:35 PM, offering information on user load, latency distribution, and system throughput:
\begin{itemize}
    \item Total Requests per Second:  The first graphic shows the number of unsuccessful requests per second (red line, labeled "Failures/s") and the rate of incoming requests (green line, called "RPS"). 
    \item Distribution of Response Times (in milliseconds) :The delay profile across three percentiles is depicted in the second chart. 
\item Multiple User Loads at Once: The number of active users (blue line) is shown in the third chart and stays at 6 for the duration of the time period. In order to ensure that observed differences in reaction time are not caused by fluctuating user volume, this steady load offers a controlled environment for evaluating the performance measurements.
\end{itemize}

\begin{figure}[!htbp]
    \centering
    \includegraphics[width=0.7\linewidth]{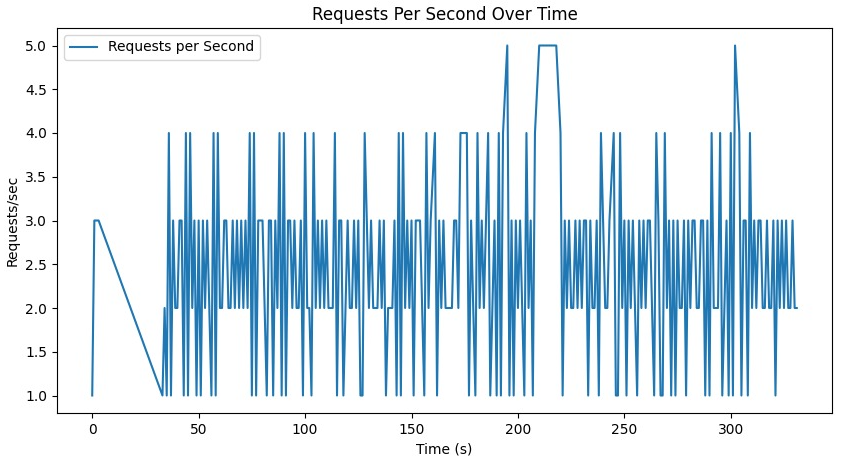}
    \caption{System Performance Metrics Over Time}
    \label{fig:sub1}
\end{figure}

\begin{figure}[!htbp]
    \centering
    \includegraphics[width=0.7\linewidth]{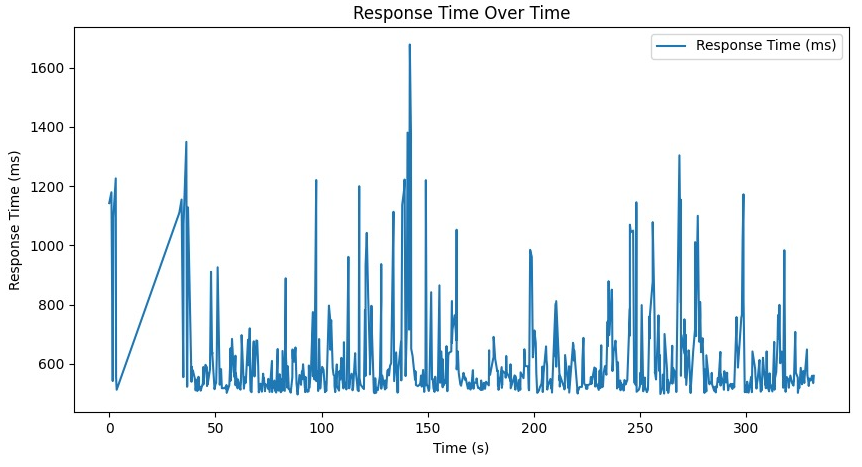}
    \caption{System Performance Metrics Over Time}
    \label{fig:sub2}
\end{figure}

\begin{figure}[!htbp]
    \centering
    \includegraphics[width=0.7\linewidth]{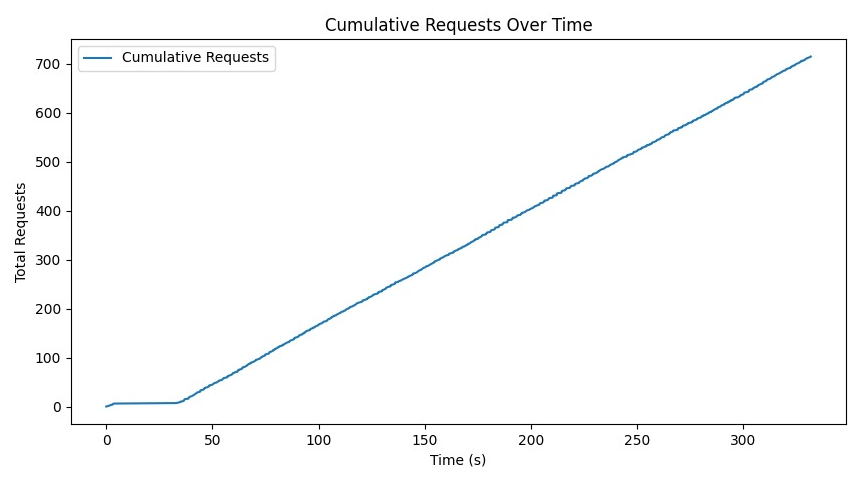}
    \caption{System Performance Metrics Over Time}
    \label{fig:sub3}
\end{figure}

 Within the first 20 seconds, the number of simulated users rapidly increases from 1 to 6, stays stable for a brief while, and then progressively decreases to 1 at the test's conclusion. In order to assess how well a system grows and recovers, load testing frequently uses this controlled increase and reduction in the number of users. Soon after the test starts, the request rate significantly increases, reaching a peak of about 4 requests per second, as seen in the Figure \ref{fig:sub1}. Following that, this rate stabilizes, showing that the system can sustain a steady load without degrading. Notably, no requests failed despite the increasing load, indicating system dependability, and the red line representing failures stays flat at 0 throughout the test. This implies that the backend infrastructure is resilient and able to preserve service integrity even in situations of traffic. The Figure \ref{fig:sub2} displays more complex performance metrics. Prior to progressively decreasing and stabilizing at lower values (about 200–300 ms), the median and 95th percentile response times first see a substantial surge, reaching approximately 1200 ms and 1500 ms, respectively. This pattern is common in cold-starting systems, where the initial queries take longer because of cache population, server warm-up, or database connection initialization. The system reaches a steady state with significantly quicker response times after these overheads are eliminated (Figure \ref{fig:sub3}). Given that response times remain constant and there are no outages during periods of high user traffic, it is proven that the solution can scale well and continue to function well even when utilized concurrently. 

%
%

\subsubsection*{ Resilience Evaluation}
To assess the resilience, stress testing was performed on the VPC that hosted the experimental setup. The aim was to emulate a volumetric application-layer attack and measure the performance of the CAHICHA-based protection strategy. The High Orbit Ion Cannon (HOIC) tool \cite{black2022overview} was chosen as the attack generator, considering that it can execute high-rate HTTP-based DoS attacks leveraging multiple concurrent threads. All the public-facing CAHICHA endpoints were set as targets in HOIC, then the tool launched a HTTP request flood that was intended to overburden the server's processing capacity.
Tcpdump was operational on the server-side during the entire length of the attack in order to record traffic patterns. The packet capture (PCAP) that was formed was then loaded into Wireshark for further analysis. A "Packets per Time" (PPT) chart was produced to show the request rate and load dynamics. In spite of the large number of requests being handled, the rate of observed TCP retransmissions, resets, and connection errors was always very low, indicating that the foundation Apache2 service and host platform remained running stability and did not crash under load. 
In addition, qualitative examination of HTTP traffic found that HOIC-forged requests demonstrated non-human patterns of automation, including consistent request headers, absence of dynamic interaction tokens, and no CAHICHA verification responses. Consequently, these requests consistently failed the CAHICHA human verification test. Therefore, the reverse proxy cut off the malicious traffic pattern at the local level and did not allow any forwarding to the backend Apache2 server on port 8080. This activity validated that CAHICHA had functioned in a successful gatekeeper mode, blocking all bot-initiated or scripted traffic while permitting legitimate, human-verified sessions to be fulfilled. It is still to be noted that this test was done on a single VPC. In bigger enterprises, CAHICHA can be deployed in a load-balanced way.

\subsubsection*{ User Accessability Test (UAT) }
We surveyed participants about their experience using CAHICHA on both mobile and web platforms, and asked whether they'd recommend it over other CAPTCHA systems like mCaptcha, reCAPTCHA, and hCaptcha. 64 people completed the CAHICHA User Accessibility Test. When we asked them to rate their mobile experience, the feedback was overwhelmingly positive—61\% gave us the highest rating of 5, while 28\% rated it a 4. Less than 2\% of participants gave low ratings of 1 or 2. The web experience told a similar story. About 59\% rated it a 5, and another 30\% gave it a 4, showing that CAHICHA performs consistently well across both platforms. When it came to recommending CAHICHA over competing systems, 64\% of respondents gave it the top recommendation score of 5, and 23\% rated it a 4. The UAT results prove that CAHICHA is an efficient solution—users clearly view it favorably in terms of accessibility, usability, and overall preference. The findings confirm CAHICHA's effectiveness as an accessible and user-friendly verification solution.


\section*{Conclusion and Future enhancements}\label{sec6}
As automation bot technology progresses, the traditional verification techniques that humans undertake as a mechanism to prevent misuse of accounts such as CAPTCHAs, voice prompts, and knowledge-based checks become less effective to AI-powered adversaries able to solve any visual puzzles, able to reproduce typing dynamics, and evade most of the measures taken today. To address this issue, we propose a hardware-interaction-based authentication technique that leverages human activities, which are inherently difficult to automate. By combining these signals along with cryptographic enforcement by tokenization and proxy validation increase usability of the verification mechanism while still being very meaningful and ultimately difficult to beat against credential stuffing, account abuse, ad fraud, and scalping attacks.
In the future, additional implementation of biometric modalities - either alongside platform input or solely, expanded implementation to IoT and edge environments, evaluation of resistance to adversarial machine learning attempts, a large scale scalability study, and placing regard for privacy in conceptual specifications will only solidify the system. We argue that hardware-supported human verification is a practical solution for the very foreseeable future and is far more appropriate than test-based peering solutions that are suggestive of the test that is Turing. Ultimately, they provide users confidence for dealing with new automated threats in the future.


\section*{Acknowledgments}
This research was conducted as part of the Research and Development (R\&D) efforts at DigitalFortress Private Limited \& Indominus Labs Private Limited. We acknowledge the company's support in facilitating this study. This work is protected and is not intended for reuse in any commercial capacity.

\bibliographystyle{IEEEtran}  
\bibliography{references}  

@misc{Pressrelease,
  author       = {{Akamai Technologies, Inc.}},
  title        = {Bots Compose 42\% of Overall Web Traffic; Nearly Two-Thirds Are Malicious},
  howpublished = {\url{https://www.akamai.com/newsroom/press-release/bots-compose-42-percent-of-web-traffic-nearly-two-thirds-are-malicious}},
  month        = jun,
  year         = 2024,
  note         = {Press release, Cambridge, MA USA – June 25, 2024}
}

@misc{Thales_2025, 
title={Building a Future we can all Trust }, 
author={{Thales Report 2025}}, url={https://www.thalesgroup.com/en/worldwide/defence-and-security/press_release/artificial-intelligence-fuels-rise-hard-detect-bots},
note    = {Available at: \url{https://www.thalesgroup.com/en/worldwide/defence-and-security/press_release/artificial-intelligence-fuels-rise-hard-detect-bots},   {$A$ccessed: 2025-05-08}
}
}

@article{debicha2023adv,
  title={Adv-Bot: Realistic adversarial botnet attacks against network intrusion detection systems},
  author={Debicha, Islam and Cochez, Benjamin and Kenaza, Tayeb and Debatty, Thibault and Dricot, Jean-Michel and Mees, Wim},
  journal={Computers \& Security},
  volume={129},
  pages={103176},
  year={2023},
  publisher={Elsevier}
}

@inproceedings{bindel2023fido2,
  title={FIDO2, CTAP 2.1, and WebAuthn 2: Provable security and post-quantum instantiation},
  author={Bindel, Nina and Cremers, Cas and Zhao, Mang},
  booktitle={IEEE Sym. Sec. Priv.},
  pages={1471--1490},
  year={2023},
  organization={IEEE}
}

@article{hoque2015botnet,
  title={Botnet in DDoS attacks: trends and challenges},
  author={Hoque, Nazrul and Bhattacharyya, Dhruba K and Kalita, 
          Jugal K},
  journal={IEEE Comm. Surv. Tut.},
  volume={17},
  number={4},
  pages={2242--2270},
  year={2015},
  publisher={IEEE}
}

@article{kwon2019security,
  title={(In-) security of cookies in HTTPS: Cookie theft by removing cookie flags},
  author={Kwon, Hyunsoo and Nam, Hyunjae and Lee, Sangtae and Hahn, Changhee and Hur, Junbeom},
  journal={IEEE Trans. Info. Foren. Sec. },
  volume={15},
  pages={1204--1215},
  year={2019},
  publisher={IEEE}
}

@article{alajmi2020password,
  title={A password-based authentication system based on the CAPTCHA AI problem},
  author={Alajmi, Masoud and Elashry, Ibrahim and El-Sayed, Hala S and Faragallah, Osama S},
  journal={IEEE Access},
  volume={8},
  pages={153914--153928},
  year={2020},
  publisher={IEEE}
}

@inproceedings{shahania2025gotta,
  title={Gotta Catch'Em All... Or Not?: How LLMs Bypass Traditional Checks \& Mimic Human Response Behavior in Web Surveys},
  author={Shahania, Saijal and Spiliopoulou, Myra and Broneske, David},
  booktitle={Proc. ACM Conf. Collect. Intell.},
  pages={113--128},
  year={2025}
}

@article{dionysiou2020sok,
  title={SoK: Machine vs. machine--A systematic classification of automated machine learning-based CAPTCHA solvers},
  author={Dionysiou, Antreas and Athanasopoulos, Elias},
  journal={Computers \& Security},
  volume={97},
  pages={101947},
  year={2020},
  publisher={Elsevier}
}

@article{wang2018delving,
  title={Delving into internet DDoS attacks by botnets: characterization and analysis},
  author={Wang, An and Chang, Wentao and Chen, Songqing and Mohaisen, Aziz},
  journal={IEEE/ACM Trans. Net.},
  volume={26},
  number={6},
  pages={2843--2855},
  year={2018},
  publisher={IEEE}
}

@article{tariq2025alert,
  title={Alert fatigue in security operations centres: Research challenges and opportunities},
  author={Tariq, Shahroz and Baruwal Chhetri, Mohan and Nepal, Surya and Paris, Cecile},
  journal={ACM Comp. Surv.},
  volume={57},
  number={9},
  pages={1--38},
  year={2025},
  publisher={ACM New York, NY}
}

@misc{Pressrelease2025, 
title={2025 Bad Bot Report}, author={Thales Cyber Security},
note={Available at: \url{https://www.imperva.com/resources/resource-library/reports/2025-bad-bot-report/},{$A$ccessed: 2025-06-18}
}
}

@article{guerar2021gotta,
  title={Gotta CAPTCHA’Em all: a survey of 20 Years of the human-or-computer Dilemma},
  author={Guerar, Meriem and Verderame, Luca and Migliardi, Mauro and Palmieri, Francesco and Merlo, Alessio},
  journal={ACM Comp. Surv.},
  volume={54},
  number={9},
  pages={1--33},
  year={2021},
  publisher={ACM New York, NY}
}

@article{wang2023experimental,
  title={An experimental investigation of text-based CAPTCHA attacks and their robustness},
  author={Wang, Ping and Gao, Haichang and Guo, Xiaoyan and Xiao, Chenxuan and Qi, Fuqi and Yan, Zheng},
  journal={ACM Comp. Surv.},
  volume={55},
  number={9},
  pages={1--38},
  year={2023},
  publisher={ACM New York, NY}
}

@article{gutub2023practicality,
  title={Practicality analysis of utilizing text-based CAPTCHA vs. graphic-based CAPTCHA authentication},
  author={Gutub, Adnan and Kheshaifaty, Nafisah},
  journal={Multi. Tools App.},
  volume={82},
  number={30},
  pages={46577--46609},
  year={2023},
  publisher={Springer}
}

@article{wang2023improving,
  title={Improving the security of audio captchas with adversarial examples},
  author={Wang, Ping and Gao, Haichang and Guo, Xiaoyan and Yuan, Zhongni and Nian, Jiawei},
  journal={IEEE Trans. Depen. Sec. Comp.},
  volume={21},
  number={2},
  pages={650--667},
  year={2023},
  publisher={IEEE}
}

@article{zhang2022counteracting,
  title={Counteracting dark Web text-based CAPTCHA with generative adversarial learning for proactive cyber threat intelligence},
  author={Zhang, Ning and Ebrahimi, Mohammadreza and Li, Weifeng and Chen, Hsinchun},
  journal={ACM Trans, Manage. Info. Sys.},
  volume={13},
  number={2},
  pages={1--21},
  year={2022},
  publisher={ACM New York, NY}
}

@article{avanesi2022m,
  title={“I’m Not a Robot,” or am I?: Micro-Labor and the Immanent Subsumption of the Social in the Human Computation of ReCAPTCHAs},
  author={Avanesi, Vino and Teurlings, Jan},
  journal={Intl. J. Comm.},
  volume={16},
  pages={19},
  year={2022}
}

@inproceedings{gangwal2025swiss,
  title={Swiss Cheese CAPTCHA: A Novel Multi-barrier Mechanism for Bot Detection},
  author={Gangwal, Ankit and Reddy, P Sahithi and Sagar, C yk},
  booktitle={Proc. ACM/SIGAPP Symp. Appl. Comp.},
  pages={1780--1789},
  year={2025}
}

@inproceedings{teoh2025captchas,
  title={Are $\{$CAPTCHAs$\}$ Still Bot-hard? Generalized Visual $\{$CAPTCHA$\}$ Solving with Agentic Vision Language Model},
  author={Teoh, Xiwen and Lin, Yun and Li, Siqi and Liu, Ruofan and Sollomoni, Avi and Harel, Yaniv and Dong, Jin Song},
  booktitle={USENIX Sec. Symp.},
  pages={3747--3766},
  year={2025}
}

@inproceedings{nayak2024experimental,
  title={Experimental security analysis of sensitive data access by browser extensions},
  author={Nayak, Asmit and Khandelwal, Rishabh and Fernandes, Earlence and Fawaz, Kassem},
  booktitle={Proc. ACM Web Conf.},
  pages={1283--1294},
  year={2024}
}

@article{golian2025integrated,
  title={Integrated graph-based testing pipeline for modern single-page applications},
  author={Golian, Nataliia and Tisheninova, Varvara},
  journal={Bulletin National Tech. Univ. " KhPI". Series: Syst. Analy. Ctrl.  Info. tech.},
  number={1 (13)},
  pages={51--59},
  year={2025}
}

@inproceedings{vasileios2021web,
  title={A web tool for analyzing FIDO2/WebAuthn Requests and Responses},
  author={Vasileios Grammatopoulos, Athanasios and Politis, Ilias and Xenakis, Christos},
  booktitle={Proc. Intl. Conf.  Avail., Relia. Sec.},
  pages={1--10},
  year={2021}
}

@article{khder2021web,
  title={Web scraping or web crawling: State of art, techniques, approaches and application.},
  author={Khder, Moaiad Ahmad},
  journal={Intl. J. Adv. Soft Comp. Appl.},
  volume={13},
  number={3},
  year={2021}
}

@inproceedings{murty2022building,
  title={Building an AI/ML based classification framework for dark web text data},
  author={Murty, Ch AS and Rana, Harmesh and Verma, Rachit and Pathak, Roshan and Rughani, Parag H},
  booktitle={Proc. Intl Conf. Comp. Comm. Net.},
  pages={93--111},
  year={2022},
  organization={Springer}
}

@inproceedings{black2022overview,
  title={An overview on detection and prevention of application layer DDoS attacks},
  author={Black, Samuel and Kim, Yoohwan},
  booktitle={Proc. Ann. Comp. Comm. Work. Conf.},
  pages={0791--0800},
  year={2022},
  organization={IEEE}
}

@inproceedings{stebila2024quantum,
  title={Quantum-safe account recovery for webauthn},
  author={Stebila, Douglas and Wilson, Spencer},
  booktitle={Proc.ACM Asia Conf. Comp. Comm. Sec},
  pages={1814--1830},
  year={2024}
}

@article{angelogianni2024many,
  title={How many FIDO protocols are needed? Analysing the technology, security and compliance},
  author={Angelogianni, Anna and Politis, Ilias and Xenakis, Christos},
  journal={ACM Comp. Surv.},
  volume={56},
  number={8},
  pages={1--51},
  year={2024},
  publisher={ACM New York, NY}
}

@inproceedings{kuchhal2023evaluating,
  title={Evaluating the security posture of real-world fido2 deployments},
  author={Kuchhal, Dhruv and Saad, Muhammad and Oest, Adam and Li, Frank},
  booktitle={Proc. ACM SIGSAC Conf. Comp. Comm. Sec},
  pages={2381--2395},
  year={2023}
}

@inproceedings{mahdad2024breaching,
  title={Breaching security keys without root: Fido2 deception attacks via overlays exploiting limited display authenticators},
  author={Mahdad, Ahmed Tanvir and Jubur, Mohammed and Saxena, Nitesh},
  booktitle={Proc. ACM SIGSAC Conf. Comp. Comm. Sec},
  pages={1686--1700},
  year={2024}
}

@article{alarnaout2022rapt,
  title={RAPT: A robust attack path tracing algorithm to mitigate SYN-flood DDoS cyberattacks},
  author={AlArnaout, Zakwan and Mostafa, Nour and Alabed, Samer and Aly, Wael Hosny Fouad and Shdefat, Ahmed},
  journal={Sensors},
  volume={23},
  number={1},
  pages={102},
  year={2022},
  publisher={MDPI}
}

@inproceedings{ghosh2024saila,
  title={Saila: Human Interface Device (HID) Injection Protection with Smart Phone based Passwordless Security},
  author={Ghosh, Anisha and Mitra, Aditya and Chakkaravarathy, S Sibi and Priya, VS Devi and Anitha, S and Babu, Rakesh Thoppaen Suresh},
  booktitle={Proc. Intl Conf. Dist. Comp. Sys.},
  pages={122--127},
  year={2024},
  organization={IEEE}
}

@article{zhang2020robust,
  title={Robust CAPTCHAs towards malicious OCR},
  author={Zhang, Jiaming and Sang, Jitao and Xu, Kaiyuan and Wu, Shangxi and Zhao, Xian and Sun, Yanfeng and Hu, Yongli and Yu, Jian},
  journal={IEEE Trans. Multi.},
  volume={23},
  pages={2575--2587},
  year={2020},
  publisher={IEEE}
}

@article{do2021methodology,
  title={A methodology for selecting hardware performance counters for supporting non-intrusive diagnostic of flood DDoS attacks on web servers},
  author={do Nascimento, Pablo Pessoa and Pereira, Paulo and Mialaret, Jr Marco and Ferreira, Isac and Maciel, Paulo},
  journal={Computers \& Security},
  volume={110},
  pages={102434},
  year={2021},
  publisher={Elsevier}
}

@inproceedings{griffioen2021scan,
  title={Scan, test, execute: Adversarial tactics in amplification DDoS attacks},
  author={Griffioen, Harm and Oosthoek, Kris and van der Knaap, Paul and Doerr, Christian},
  booktitle={Proc. ACM SIGSAC Conf. Comp. Comm. Sec.},
  pages={940--954},
  year={2021}
}

\end{document}